\documentclass[preprint,superscriptaddress,nofootinbib,notitlepage, double-spaced, showpacs,floatfix,longbibliography]{revtex4-2}
\usepackage{epsfig}
\usepackage{amsthm}
\usepackage{amsfonts}
\usepackage{float}
\usepackage{amsmath,amssymb}
\usepackage{enumerate}
\usepackage{color}
\usepackage{fancyvrb}
\usepackage[colorlinks = true,
linkcolor = blue,
urlcolor  = blue,
citecolor = red,
anchorcolor = green]{hyperref}
\usepackage{graphicx}
\usepackage{dcolumn}
\usepackage{bm}
\usepackage{hyperref}
\usepackage[normalem]{ulem}
\usepackage{xcolor}
\graphicspath{{./figs/}}
\usepackage[caption=false]{subfig}
\usepackage[percent]{overpic}
\renewcommand{\d}{\mathrm{d}}

\begin{document}
	
	\title{Bayesian constraints on the transport coefficients $\eta/s$ and $\zeta/s$ from spin polarization in relativisitic heavy-ion collisions}
	
	\author{Sushant K. Singh}
	\email{sushant7557@gmail.com}
	\affiliation{Dipartimento di Fisica, Universit\`a di Firenze, via G. Sansone 1,
		50019 Sesto Fiorentino, Italy}
	\affiliation{Variable Energy Cyclotron Centre, 1/AF Bidhan Nagar, Kolkata- 700064, India}
	
	\author{Eduardo Grossi}
	\email{eduardo.grossi@unifi.it}
	\affiliation{Dipartimento di Fisica, Universit\`a di Firenze, via G. Sansone 1,
		50019 Sesto Fiorentino, Italy}
	\affiliation{INFN Sezione di Firenze, via G. Sansone 1, 50019 Sesto Fiorentino, Italy}
	
	\author{Francesco Becattini}
	\email{becattini@fi.infn.it}
	\affiliation{Dipartimento di Fisica, Universit\`a di Firenze, via G. Sansone 1,
		50019 Sesto Fiorentino, Italy}	
	\affiliation{INFN Sezione di Firenze, via G. Sansone 1, 50019 Sesto Fiorentino, Italy}
	
	\begin{abstract}
		Bayesian analyses in the context of relativistic heavy-ion collisions have so far relied almost exclusively on bulk hadronic observables constructed from momentum degrees of freedom to constrain the transport properties of the quark-gluon plasma. In this work, we perform the Bayesian inference after incorporating the longitudinal spin polarization of $\Lambda$ hyperons alongside conventional bulk measurements in Pb+Pb collisions at $\sqrt{s_{NN}}=5.02$ TeV to constrain the shear and bulk viscosity to entropy density ratios, $\eta/s$ and $\zeta/s$. We demonstrate that the inclusion of spin polarization, which provides complementary sensitivity to the space-time structure and vorticity of the medium, shifts the posterior distribution of $\zeta/s$ toward larger values, although current uncertainties do not allow a statistically significant separation at the 68\% credibility level. Nevertheless, the results establish spin polarization as a valuable probe in quantitative studies of QGP transport properties and indicate that it should be incorporated in comprehensive and systematically constrained Bayesian extractions of the medium's dynamical parameters.
	\end{abstract}

	\maketitle
	
	
	\newpage
	
	\section{\label{intro} Introduction}
	
	The remarkable success of relativistic hydrodynamics in describing collective flow observables in heavy-ion collisions has led to the conclusion that the quark–gluon plasma (QGP) behaves as an almost perfect fluid~\cite{Gale:2013ee,Heinz:2013th,Niemi:2016ee}. In particular, comparisons of hydrodynamic simulations with elliptic flow measurements indicate that the specific shear viscosity, $\eta/s$, is very small and close to the conjectured lower bound~\cite{Kovtun:2005vi}, making the QGP the most perfect fluid observed in nature. Despite this phenomenological success, the transport coefficients of QCD matter, including both the shear viscosity $\eta$ and the bulk viscosity $\zeta$, have not yet been computed reliably from first principles in the temperature regime relevant for heavy-ion collisions. The strongly coupled and non-perturbative nature of QCD near the confinement scale renders perturbative approaches inadequate, while lattice calculations of transport coefficients remain notoriously challenging~\cite{Astrakhantsev:2017nrs, Altenkort:2023vig}. Nevertheless, these coefficients play a central role in governing the space-time evolution of the plasma and are therefore essential for a quantitative characterization of the medium.
	
	To address this difficulty, the community has adopted a top-down strategy in which transport coefficients are inferred by systematically comparing model calculations with experimental data. In recent years, Bayesian inference techniques have emerged as a powerful framework for this purpose~\cite{Bernhard:2016amu,Bernhard:2019bmu,Everett:2021dsh,Parkkila:2021dbj,Trajectum2021,Vermunt:2023zsk,Jahan:2024bar,rzml-rjxz}, enabling a statistically rigorous extraction of these coefficients with quantified uncertainties from a broad set of observables measured at RHIC and the LHC. These analyses have primarily relied on hadronic observables such as multiplicities, mean transverse momenta and anisotropic flow coefficients, which are sensitive to the viscous response of the medium. Such global Bayesian calibrations have significantly improved our understanding of the temperature dependence of $\eta/s$ and $\zeta/s$, yet they have so far been restricted to momentum-space observables and have not incorporated spin-related measurements.
	
	In parallel with its near-perfect fluidity, the QGP has also been identified as the most vortical fluid ever observed, as evidenced by the global spin polarization of $\Lambda$ hyperons measured in heavy-ion collisions~\cite{STAR:2017ckg}. The description of spin observables such 
	as global polarization and spin alignment has stimulated intense theoretical activity from different perspectives 
	(see reviews \cite{Becattini:2020ngo,Becattini:2022zvf,Dong:2024yuq,Becattini:2024uha} and references therein).
	
	Recent studies have shown that $\Lambda$ polarization in heavy-ion collisions is sensitive to the medium's transport properties~\cite{Singh2023hdo}, with the longitudinal polarization component, $P_z$, being particularly sensitive to the bulk viscosity and initial flow structure at LHC energies~\cite{Palermo:2024tza}. Moreover, extensions of kinetic theory incorporating spin degrees of freedom indicate that spin polarization itself can modify effective transport coefficients, including bulk and shear viscosity, thereby influencing the thermodynamic evolution of the fluid~\cite{Wei:prc2025}. These developments motivate the inclusion of spin polarization measurements in a unified Bayesian analysis of QGP transport coefficients, as their demonstrated sensitivity to bulk viscosity can provide additional and independent constraints beyond conventional measurements constructed from momentum degrees of freedom. In this work, we perform a systematic Bayesian study of $\eta/s$ and $\zeta/s$ by incorporating longitudinal polarization data as an additional constraint alongside traditional hadronic observables. Because the computation of spin polarization requires full (3+1)-dimensional hydrodynamic simulations and a realistic event-by-event treatment is computationally demanding, we carry out our analysis using event-averaged initial profiles. A detailed event-by-event treatment including pre-equilibrium dynamics, while important, is left for future work. To make the parameter estimation tractable, we construct a Gaussian process emulator trained on a carefully designed set of hydrodynamic calculations, which is subsequently employed within a Markov Chain Monte Carlo (MCMC) framework to extract posterior distributions of the transport coefficients. This study provides the first Bayesian constraints on QGP viscosities that explicitly incorporate spin polarization observables, thereby opening a new avenue for probing the dissipative properties of strongly interacting matter.
	
	{\it Notation:} We work in natural units where $\hbar=k_B=c=1$. We adopt the mostly negative metric convention, $g_{\mu\nu}=\text{diag}(1,-1,-1,-1)$. Our code is written in Milne coordinates, $(\tau,x,y,\eta_s)$, expressed in terms of Cartesian coordinates as defined as $\tau=\sqrt{t^2-z^2}$ and $\eta_s =\frac{1}{2}\text{ln}\left(\frac{t+z}{t-z}\right)$. The inner product of two four-vectors is denoted by $A\cdot B = g_{\mu\nu}A^\mu B^\nu$. The fluid velocity is denoted by $u^\mu$ such that $u\cdot u = 1$.
	The covariant derivative of a quantity $A$ is denoted by $\nabla_\mu A$.
	The comoving derivative $u\cdot \partial$ of a quantity $A$ is denoted by $\dot{A}=u\cdot \nabla A$. The derivative orthogonal to the 4-velocity is written as $\nabla_{\perp}^\mu =\Delta^{\mu\nu}\nabla_\nu$. For the Levi-Cività tensor $\epsilon^{\mu\nu\alpha\beta}$, we follow the sign convention $\epsilon^{0123} =-\epsilon_{0123} = +1$.
	
	\section{Description of the model} 
	
	The hydrodynamic evolution is governed by energy-momentum and net baryon number conservation, and which are expressed through the following continuity equations:
	\begin{align}
		\nabla_\mu T^{\mu\nu} (x) &= 0\;,
		\label{eq:T_cons}\\
		\nabla_\mu N^\mu (x) &= 0\;.
		\label{eq:N_cons}
	\end{align}
	Here $T^{\mu\nu}$ is the energy-momentum tensor, and $N^{\mu}$ is the (net baryon) charge current and the $\nabla_\mu$ is the covariant derivative of the spacetime. We choose to work in the Landau frame, defined by
	\begin{equation}
		\label{eq:Lframe}
		T^{\mu}{}_{\nu}  u^\nu= \varepsilon\, u^\mu\;,
	\end{equation}
	with  $\varepsilon$ the local energy density and $u^{\nu}$ the fluid velocity. 
	With this choice, the energy-momentum tensor and the net baryon current are generically written as
	\begin{align}
		\label{eq:T_curr} 
		T^{\mu\nu} &= \varepsilon u^\mu u^\nu - (P+\Pi) \Delta^{\mu\nu} + \pi^{\mu\nu}\;,\\
		\label{eq:N_curr} 
		N^{\mu} &=  n_B\,u^\mu + V^\mu\;.
	\end{align}
	Here $\varepsilon$ is the energy density, $n_B$ is the net-baryon number density, $P$ is the thermodynamic equilibrium pressure, $\pi^{\mu\nu}$ is the shear-stress tensor, $\Pi$ is the bulk viscous pressure, and $V^\mu$ is the baryon diffusion current, and $\Delta_{\mu\nu} =g_{\mu\nu}-u_{\mu}u_\nu$ is the projector transverse to the fluid velocity $u_\mu$. In this study, we consider $V^\mu=0$ for simplicity. The equations of motion for $\pi^{\mu\nu}$ and $\Pi$ are taken as~\cite{Denicol:2012cn,Denicol:2015transcoeff}
	\begin{align}
		\dot{\Pi}&=\frac{\Pi_{\rm N S}-\Pi}{\tau_{\Pi}}-\frac{\delta_{\Pi \Pi}}{\tau_{\Pi}} \Pi\, \theta+\frac{\lambda_{\Pi \pi}}{\tau_{\Pi}} \pi^{\alpha\beta} \sigma_{\alpha\beta} \label{eq:Pi}\;,\\
		\dot{\pi}^{\langle\alpha\beta\rangle}&=\frac{\pi_{\rm N S}^{\alpha \beta}-\pi^{\alpha\beta}}{\tau_\pi}
		-\frac{\delta_{\pi \pi}}{\tau_\pi} \pi^{\alpha \beta} \theta+\frac{\lambda_{\pi \Pi}}{\tau_\pi} \Pi \,\sigma^{\alpha\beta} -\frac{\tau_{\pi \pi}}{\tau_\pi} \pi_\gamma^{\langle\alpha} \sigma^{\beta\rangle \gamma}+\frac{\varphi_7}{\tau_\pi} \pi_\gamma^{\langle\alpha} \pi^{\beta\rangle \gamma}\;\label{eq:pi}.
	\end{align}
	where $\pi_{\rm NS}^{\mu\nu}$ and $\Pi_{\rm NS}$ are corresponding Navier-Stokes limit given by
	\begin{equation*}
		\pi_{\rm NS}^{\mu\nu}= 2\eta\,  \Delta_{\gamma \delta}^{\mu\nu} \nabla^\gamma u^\delta = 2 \eta\sigma^{\mu\nu} \quad , \quad \Pi_{\rm NS} = - \zeta \nabla_\gamma u^\gamma = - \zeta \theta. 
	\end{equation*}
	Here $\eta>0$ and $\zeta >0$ denote the shear and bulk viscosities, respectively. These are treated as temperature-dependent quantities with no $\mu_B$ dependence. We parameterize these transport coefficients using the following functional forms~\cite{Trajectum2021}
	\begin{equation}
		\frac{\eta}{s} = \left\{ \begin{array}{cc}
			(\eta/s)_{\text{min}} + (\eta/s)_{\text{slope}}(T-T_\eta)\left(\frac{T}{T_\eta}\right)^{(\eta/s)_{\text{curv}}} & T\ge T_\eta,\\
			0.06 & T < T_\eta
		\end{array}\right.
		\label{eq:etaS}
	\end{equation}
	with $T_\eta = 154$ MeV, and
	\begin{equation}
		\frac{\zeta}{s} = \frac{(\zeta/s)_{\text{max}}}{1+\left[\frac{T-(\zeta/s)_{\text{T0}}}{(\zeta/s)_{\text{width}}}\right]^2}.
		\label{eq:zetaS}
	\end{equation}
	In particular, we introduce three free parameters each to characterize $\eta/s(T)$ and $\zeta/s(T)$. The remaining second-order transport coefficients entering the equations ~\eqref{eq:Pi}-\eqref{eq:pi} are not treated as independent parameters. Instead, they are fixed using expressions derived in kinetic theory~\cite{Denicol:2015transcoeff,Transcoeff2014_p7}, but evaluated consistently with the chosen equation of state (discussed below). This approach reduces the dimensionality of the parameter space while retaining a physically motivated description of the system. 
	The remaining transport coefficients are expressed below:
	\begin{align}
		&\tau_\pi = \frac{5\eta}{\varepsilon+P} \;,\quad \tau_\Pi = \frac{\zeta}{15\left(\frac{1}{3}-c_s^2\right)^2(\varepsilon+P)}\;,\nonumber\\
		&\frac{\delta_{\Pi\Pi}}{\tau_\Pi} = \frac{2}{3}	\;, \quad  \frac{\lambda_{\Pi\pi}}{\tau_\Pi} = \frac{8}{5}\left(\frac{1}{3}-c_s^2\right)\;,\quad \frac{\delta_{\pi\pi}}{\tau_\pi} = \frac{4}{3}\;,\nonumber\\
		& \frac{\lambda_{\pi\Pi}}{\tau_\pi} = \frac{6}{5}\;,\quad \frac{\tau_{\pi\pi}}{\tau_\pi} = \frac{10}{7}\;,\quad \frac{\varphi_7}{\tau_\pi} = \frac{9}{70 P \tau_\pi}\;.\label{eq:coef}
	\end{align}
	We employ the lattice-QCD-based equation of state NEOS-BQS~\cite{Monnai2019eos,HotQCD:2014kol,HotQCD:2012fhj,Ding:2015fca,Bazavov:2017dus,eosweblink} to close the hydrodynamic equations. The square of the speed of sound is computed from the equation of state as follows
	\begin{equation}
		c_s^2=\left.\frac{\partial P}{\partial \varepsilon}\right|_{n_B}
		+\left. \frac{n_B}{\varepsilon+P} \frac{\partial P}{\partial n_B}\right|_{\varepsilon}\;.
	\end{equation}
	
	As mentioned above, we do not explicitly model the pre-equilibrium stage in this work. Instead, we assume that the system has locally thermalized at a proper time $\tau_0$, which is treated as a free parameter, after which a hydrodynamic description becomes applicable. The initial energy-momentum tensor at $\tau_0$ is constructed using the model described in Refs~\cite{Shen:2020jwv,Shen:2021lambda}, and is given by
	\begin{align}
		T^{\tau\tau}(\tau_0;x,y,\eta_s) &= \varepsilon(\tau_0;x,y,\eta_s)\cosh (f Y), \\
		T^{\tau x}(\tau_0;x,y,\eta_s) &=T^{\tau y}(\tau_0;x,y,\eta_s)=0, \\
		T^{\tau\eta_s}(\tau_0;x,y,\eta_s) &= \frac{1}{\tau_0}\varepsilon(\tau_0;x,y,\eta_s)\sinh (f Y),
	\end{align}
	where $f$ is a parameter controlling the fraction of longitudinal momentum converted into fluid velocity and $Y$ is defined as
	$$Y(x,y) = \tanh^{-1}\left[ \frac{T_A-T_B}{T_A+T_B}\tanh (y_{\text{beam}})\right] \quad \text{with } \quad y_{\text{beam}}  = \cosh^{-1}\left(\frac{\sqrt{s_{\rm NN}}}{2m_N}\right). $$
	In the above expressions, $T_A$ and $T_B$ are nucleus thickness functions computed from the positions of participant nucleons using the following expression
	\begin{equation}
		T_{A(B)}(x,y) = \sum_{i\in \text{participants}}\frac{1}{2\pi w^2}\exp\left[-\frac{1}{2w^2}\left\{(x-x_i)^2+(y-y_i)^2\right\}\right],
		\label{eq:thickness}
	\end{equation}
	where $(x_i,y_i)$ denotes the transverse coordinates of the $i^{\text{th}}$ nucleon and $w$ is the width of the nucleon.
	
	For symmetric collisions, the initial energy density is given by
	\begin{equation}
		\varepsilon (\tau_0;x,y,\eta_s) = \varepsilon_\perp (x,y) \, F(\eta_s),
	\end{equation}
	with $F(\eta_s)$ defined as~\cite{Shen:2020jwv}
	\begin{align}
		F(\eta_s) &= \exp\Bigg[ -\frac{(|\eta_s -(1-f)Y|-\eta_0)^2}{2\sigma_\eta^2} \Theta(|\eta_s -(1-f)Y|-\eta_0)\Bigg].
	\end{align}
	Here $\Theta(x)$ denotes the Heaviside function. $\varepsilon_\perp(x,y)$ is obtained through the matching condition at the initial time, which results in an expression $\varepsilon_\perp(x,y) =\mathcal{N}_\varepsilon M(x,y)$, with $\mathcal{N}_\varepsilon$ expressed in terms of $\eta_0$ and $\sigma_\eta$. The choice of initial energy-momentum tensor results in zero transverse velocity, and the $\eta_s$-component of the velocity is determined numerically by solving the implicit equation
	$$v^{\eta_s}(\tau_0;x,y,\eta_s) = \frac{T^{\tau\eta_s}(\tau_0;x,y,\eta_s)}{T^{\tau\tau}(\tau_0;x,y,\eta_s)+P_{\rm eq}(\varepsilon,n)}.$$
	For the initial baryon density, we consider the following profile 
	\begin{equation}
		n_B (\tau_0;x,y,\eta_s) = \mathcal{N}_n\left[ g_A(\eta_s)T_A(x,y) + g_B(\eta_s)T_B(x,y)\right],
	\end{equation}
	where $g_A(\eta_s)$ and $g_B(\eta_s)$ are given by~\cite{Denicol:2018wdp}:
	\begin{align}
		g_A(\eta_s) &=\Theta (\eta_s-\eta_{B,0})
		\exp\left[-\frac{(\eta_s-\eta_{B,0})^2}{2\sigma _{B,{\texttt{out}}}^2}\right]+\Theta (\eta_{B,0}-\eta_s)\exp \left[-\frac{(\eta_s-\eta_{B,0})^2}{2\sigma_{B,{\texttt{in}}}^2}\right],
	\end{align}
	\begin{align}
		g_B(\eta_s) &= \Theta (\eta_s +\eta_{B,0})\exp \left[  -\frac{(\eta_s+\eta_{B,0})^2}{2\sigma _{B,{\texttt{in}}}^2}\right] + \Theta (-\eta_{B,0}-\eta_s)\exp \left[  -\frac{(\eta_s+\eta_{B,0})^2}{2\sigma _{B,{\texttt{out}}}^2}\right].
	\end{align}
	The normalization constant $\mathcal{N}_n$ is chosen by demanding that
	\begin{equation}
		\int \, \tau_0 \, dx\, dy\, d\eta_s \ n_B (\tau_0;x,y,\eta_s) = N_{\text{part}} 
	\end{equation}
	where $N_{\text{part}}$ denotes the total number of participants. The above condition gives
	\begin{equation}
		\mathcal{N}_n = \frac{1}{\tau_0}\sqrt{\frac{2}{\pi}}\frac{1}{\sigma _{B,{\texttt{in}}}+\sigma _{B,{\texttt{out}}}}.
	\end{equation}
	We fix the parameter $\sigma _{B,{\texttt{out}}}=0.1$. For the net charge density profile, we consider
	$$n_Q(x,y) \approx 0.4 \ n_B(x,y)$$
	The hydrodynamic evolution is stopped at the particlization (or switching) hypersurface, defined by a constant energy density $\varepsilon_{\text{sw}}$. This hypersurface is obtained using the \texttt{CORNELIUS} code~\cite{Huovinen2012}. At this 
	hypersurface, we use a hadron sampler~\cite{Karpenko:2015xea,Schafer2022,samplerweblink} to generate particles from fluid 
	elements. The resulting particle set serves as input to the SMASH transport model~\cite{SMASH2016prc,wergieluk_2024_10707746} 
	for subsequent hadron interactions and decays. 
	
	We compute the spin polarization of $\Lambda$-hyperons on the switching hypersurface, including the contributions from thermal vorticity,
	$$\varpi_{\nu\rho}=\frac{1}{2}(\nabla_\rho \beta_\nu-\nabla_\nu\beta_\rho)$$
	and thermal shear, 
	$$\xi_{\nu\rho}=\frac{1}{2}(\nabla_\rho \beta_\nu+\nabla_\nu\beta_\rho)$$
	following the prescription of Refs.~\cite{Becattini:2015ska,Becattini:2021suc} \footnote{It should be noted that a 
		different form of the shear contribution was proposed in refs. \cite{Liu:2021uhn} and that, lately, an upgraded formula 
		was calculated in ref. \cite{Sheng:2025cjk}. For the purpose of testing the response of longitudinal polarization 
		to bulk viscosity, which is the main goal of this work, we confine ourselves to the form in \cite{Becattini:2021suc} 
		because it has been successfully tested against data in ref. \cite{Palermo:2024tza}.}. The spin four-vector is written as
	\begin{equation}
		\label{eq:bbp}
		S^\mu(p) = S^\mu_{\varpi}(p) + S^\mu_{\xi}(p)\;,
	\end{equation}
	with 
	\begin{subequations}
		\label{eqs:components_S_BBP}
		\begin{align}
			S_{\varpi}^\mu(p) &= -\frac{1}{8m_\Lambda}\epsilon^{\mu\nu\rho \sigma}p_\sigma \frac{\int \d\Sigma\cdot p \ f_0(1-f_0) \varpi_{\nu\rho}}{\int \d\Sigma\cdot p \ f_0}\;,\\
			S^\mu_{\xi}(p) &= -\frac{\epsilon^{\mu\nu\rho \sigma}}{4m_\Lambda} \frac{p_\sigma p^\lambda}{p\cdot \hat{t}}\frac{\int \d\Sigma\cdot p \ f_0(1-f_0) \hat{t}_\nu \xi_{\lambda\rho}}{\int \d\Sigma\cdot p \ f_0}\;.
		\end{align}
	\end{subequations}
	In the above equation, $f_0$ is the relativistic Fermi-Dirac distribution with four-temperature $\beta(x)$ and the
	appropriate combination of chemical potentials for the $\Lambda,\bar\Lambda$ hyperons; in the shear contribution, $\hat{t}^\mu$ is the time unit vector $\hat{t}^\mu=(1,0,0,0)$ in the centre-of-mass collision frame. At very high collision energies, the baryon chemical potential is very small and satisfies $\mu_B\approx0$, such that energy-density becomes primarily a function of temperature. In this limit, the constant energy density switching hypersurface can be approximated by a constant temperature hypersurface after neglecting the gradients of temperature on the hypersurface. This allows the temperature to be factored out of the surface integrals in the density operator. This is known as the isothermal approximation \cite{Becattini:2021suc}. Under this approximation, defining $\Xi^{\mu\nu}=\frac12 \partial^{(\mu}u^{\nu)}$ and $\omega^{\mu\nu}= \frac12 \partial^{[\mu}u^{\nu]}$, Eqs. \eqref{eqs:components_S_BBP} reduce to
	\begin{subequations}
		\begin{align}
			S_{\varpi}^{\mu}(p) &= -\frac{1}{8m_\Lambda}\epsilon^{\mu\nu\rho \sigma}p_\sigma \frac{\int \d\Sigma\cdot p \ f_0(1-f_0) \omega_{\nu\rho}}{T\int \d\Sigma\cdot p \ f_0}\;,\\
			S^{\mu}_{\xi}(p) &= -\frac{\epsilon^{\mu\nu\rho \sigma}}{4m_\Lambda} \frac{p_\sigma p^\lambda}{p\cdot \hat{t}}\frac{\int \d\Sigma\cdot p \ f_0 (1-f_0) \hat{t}_\nu \Xi_{\lambda\rho}}{T\int \d\Sigma\cdot p \ f_0}\;.
		\end{align}
	\end{subequations}
	The spin vector in the $\Lambda$ rest frame, $S^{*\mu} = (0,\textbf{S}^*)$ is obtained from the laboratory frame spin vector $S^\mu = (S^0,\textbf{S})$ as follows:
	\begin{equation*}
		\textbf{S}^* = \textbf{S}-\frac{\textbf{p} \cdot \textbf{S}}{E(E+m_\Lambda)}\textbf{p}\;.
	\end{equation*}
	where $E=p^0$ is the $\Lambda$ energy in the laboratory frame. Finally, the spin polarization vector is given by $\textbf{P}=2\textbf{S}^*$. For simplicity and to reduce computational cost, feed-down contributions from resonance decays are neglected in the present analysis. This is a reasonable approximation, as it has been shown that the inclusion of feed-down corrections does not lead to significant changes in the resulting $\Lambda$ polarization relative to the case of entirely primary production~\cite{Palermo:2024tza}.

	\section{Model Parameters}
	
	A full (3+1)-dimensional event-by-event simulation of Pb+Pb collisions at $\sqrt{s_{NN}}=5.02$ TeV is computationally demanding, particularly for central collisions where the fireball lifetime is comparatively long and particle multiplicities are large. The computational cost is further amplified by the particlization (sampler) stage and the subsequent hadronic transport evolution. To render the Bayesian analysis feasible while maintaining a physically well-motivated framework, we adopt a set of simplifications. We neglect pre-equilibrium dynamics and initialize hydrodynamics at a fixed proper time $\tau_0$. The initial condition model contains six ($f$, $\sigma_\eta$, $\eta_0$, $\eta_{B,0}$, $\sigma_{B,\text{in}}$, $\sigma_{B,\text{out}}$) parameters and is constructed from thickness functions in which the smoothing of local hotspots is regulated by the nucleon width $w$. For each parameter set, we generate multiple fluctuating initial configurations to obtain the energy-momentum tensor $T^{\mu\nu}$ for individual events, and subsequently perform event averaging to produce a smooth $T^{\mu\nu}$ profile for hydrodynamic evolution. The transport sector includes six parameters describing the temperature dependence of $\eta/s$ and $\zeta/s$ (three each), together with the switching energy density, $\varepsilon_{\text{sw}}$, defining the transition from hydrodynamics to the hadronic afterburner, resulting in a total of fifteen model parameters. 
	
	\begin{figure*}[t]
		\centering
		\includegraphics{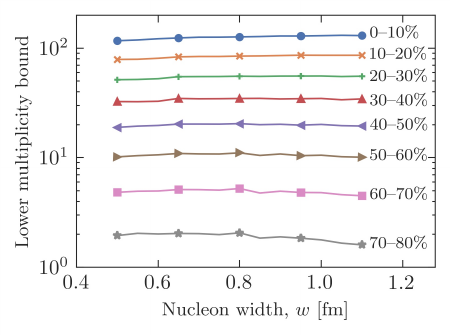}
		\caption{Scaled multiplicity vs nucleon width}
		\label{fig:centvswidth}
	\end{figure*}
	
	We find that two parameters, $w$ and $\sigma_{B,\text{out}}$, have negligible impact on the calibrated observables within reasonable ranges; these are therefore fixed to $w=0.9$ and $\sigma_{B,\text{out}}=0.2$, reducing the effective dimensionality for the analysis to thirteen parameters. For example, in Fig.~\ref{fig:centvswidth}, we show the dependence of the lower bound of the scaled multiplicity range on $w$ for various centrality classes. The bound remains approximately constant over a wide range of $w$, demonstrating that the centrality selection is largely insensitive to this parameter. We adopt uniform prior distributions for all model parameters within their respective ranges, reflecting minimal prior knowledge. The prior ranges for the parameters are summarized in Table~\ref{tab::param_prior_map}. 
		
	\section{\label{sec:expdata}Experimental data}
	We consider nine sets of experimental data across seven centrality classes: 0-10\%, 10-20\%, 20-30\%, 30-40\%, 40-50\%, 50-60\%, and 60-70\%. The observables included in the Bayesian analysis are listed below:
	\begin{enumerate}[(i)]
		\item Charged particle pseudorapidity density, $\frac{dN_{\text{ch}}}{d\eta}$ from Refs.~\cite{ExptData_dNdeta_5020, ExptData_identified_dNdy_5020},
		\item Identified-particle rapidity densities, $\left.\frac{dN}{dy}\right|_{\pi^++\pi^-}$, $\left.\frac{dN}{dy}\right|_{K^++K^-}$, and $\left.\frac{dN}{dy}\right|_{p+\bar{p}}$ from Ref.~\cite{ExptData_identified_dNdy_5020}, 
		\item Mean transverse momentum of identified particles $\langle p_T\rangle_{\pi^++\pi^-}$, $\langle p_T\rangle_{K^++K^-}$, and $\langle p_T\rangle_{p+\bar{p}}$ from Ref.~\cite{ExptData_identified_dNdy_5020},
		\item Elliptic flow coefficient $v_2$ of charged particles from Ref.~\cite{ExptData_v2_5020},
		\item Longitudinal spin polarization $P_z$ of $\Lambda$ hyperons from Ref.~\cite{ALICE2022:ppz}.
	\end{enumerate}
	The experiemental data listed in items (i)-(iv) are not available in the 0-10\% centrality class but are reported separately for the 0-5\% and 5-10\% centrality bins. We therefore merge these two bins to construct the corresponding 0-10\% observables following the procedure described below. 
		
	The combined experimental values of $\frac{dN_{\mathrm{ch}}}{d\eta}$ and $\frac{dN}{dy}$ are taken as the arithmetic mean of the corresponding values in the two centrality classes, while the associated uncertainty is taken to be the larger of the two quoted uncertainties. For the mean tranverse momentum, the value in the merged centrality class is computed as
	$$\langle p_T\rangle = \frac{\left(\frac{dN}{dy}\right)_{C_1}\langle p_T\rangle_1+\left(\frac{dN}{dy}\right)_{C_2}\langle p_T\rangle_2}{\left(\frac{dN}{dy}\right)_{C_1}+\left(\frac{dN}{dy}\right)_{C_2}}$$
	where $C_1$ and $C_2$ denote the 0-5\% and 5-10\% centrality classes, respectively. The corresponding uncertainty is taken as the larger of the two quoted uncertainties.
	
	For the elliptic flow coefficient $v_2$, an additional step is required. Our hydrodynamic simulations employ event-averaged smooth initial profiles and therefore predict the mean elliptic flow $\langle v_2\rangle$, while experimental measurements include contributions from event-by-event flow fluctuations. To make the experimental data compatible with our model, we follow Ref.~\cite{Ollitrault2009:fluct} and characterize the magnitude of flow fluctuations by a variance $\sigma^2$. The elliptic flow coefficients measured using the two- and four-particle cumulant methods satisfy 
	$$v_2\{2\}^2 = \langle v_2\rangle^2 +\sigma^2\quad , \quad v_2\{4\}^2 \approx \langle v_2\rangle^2 -\sigma^2$$
	From these relations, the mean elliptic flow is obtained as
	$$\langle v_2\rangle = \sqrt{\frac{v_2\{2\}^2+v_2\{4\}^2}{2}}$$
	Measurements of $v_2\{2\}$ and $v_2\{4\}$ are available for Pb+Pb collisions at $\sqrt{s_{NN}}=5.02$ TeV~\cite{ExptData_v2_5020}. We use these measurements to extract $\langle v_2\rangle$ in each experimental centrality bin, with uncertainties propageted in quadrature. The values in the 0-5\% and 5-10\% classes are then merged using the same procedure as for the mean transverse momentum.
	
	\section{Computational details}
	
	The initial condition model employed in this work satisfies the scaling $\varepsilon_\perp\sim \sqrt{T_AT_B}$ at high collision energies. The T$_\textsc{R}$ENTo model~\cite{Trento:2015sdg} provides this behavior for the reduced thickness parameter $p=0$. We therefore use the T$_\textsc{R}$ENTo code~\cite{trentoweblink} to construct the initial conditions and to perform centrality selection.
	
	We generate $10^5$ minimum-bias events in the impact-parameter range 0-20 fm using the following T$_\textsc{R}$ENTo configuration : 
	\begin{center}
		\begin{Verbatim}[frame=single, baselinestretch=0.8, fontsize=\small]
			cross-section = 7.0
			normalization = 1.0
			reduced-thickness = 0.0
			fluctuation = 1.0
			nucleon-min-dist = 0.0
			nucleon-width = 0.9
			b-min = 0.0
			b-max = 20.0
			grid-max = 10.0
			grid-step = 0.2
		\end{Verbatim}
	\end{center}
	For each event, we store the transverse coordinates of the participating nucleons and compute the integrated reduced thickness \(\int d^2x_\perp \sqrt{T_A T_B}\). We assume the final-state charged-particle multiplicity to be proportional to this quantity and sort the events in descending order of the integrated reduced thickness. Since centrality classes are defined by percentiles of the multiplicity distribution, the overall normalization (set here to unity) does not affect the centrality determination.
	
	Centrality classes are defined by selecting fixed percentiles of the sorted event ensemble in intervals of 10\%. For each centrality class, we select 5000 events from the minimum-bias sample. For each selected event, the nuclear thickness functions are constructed using Eq.~(\ref{eq:thickness}), and the
	corresponding energy--momentum tensor \(T^{\mu\nu}\) is evaluated. The initial condition for the hydrodynamic evolution is then obtained by averaging over all events in a given centrality class, yielding a smooth, event-averaged energy--
	momentum tensor \(\langle T^{\mu\nu} \rangle\).
	
	The viscous hydrodynamics is solved numerically in full (3+1) dimensions using the code developed in Refs.~\cite{Singh:2024cub,Singh2023hdo,Singh2023qcp}. The code, written in \texttt{Fortran}, employs the Godunov-type relativistic Harten-Lax-van Leer-Einfeldt (HLLE) approximate Riemann solver to compute numerical fluxes at fluid cell interfaces, as described in Refs.~\cite{RISCHKE1995346,Karpenko:2013wva}. Although inspired by the publicly available \texttt{vHLLE} framework~\cite{Karpenko:2013wva,vhlleweblink}, the code includes several extensions, such as multi-threaded parallelization via \texttt{OpenMP}~\cite{openmpweblink} and the evolution of spin degrees of freedom within a spin-hydrodynamic framework~\cite{Sapna:2025}. For the present analysis, only the background viscous hydrodynamic evolution is employed.
	
	Our analysis is based on a finite ensemble of model evaluations performed at design points in the model parameter space. The design matrix \textbf{X}$_\text{all}$ is constructed using maximin Latin hypercube sampling (LHS), which provides an approximately space-filling coverage of the multidimensional parameter space~\cite{Bernhard:2016amu,Trajectum2021,Stein01051987}. The corresponding simulation outputs \textbf{Y}$_\text{all}$ contain the predicted observables for Pb+Pb collisions at $\sqrt{s_{NN}}$=5.02 TeV. The observable vector entering the Bayesian analysis has dimension $n_{\text{obs}}=63$, corresponding to nine observables evaluated in seven centrality classes (refer section~\ref{sec:expdata}).
	
\begin{figure*}[t]
		\centering
		\includegraphics{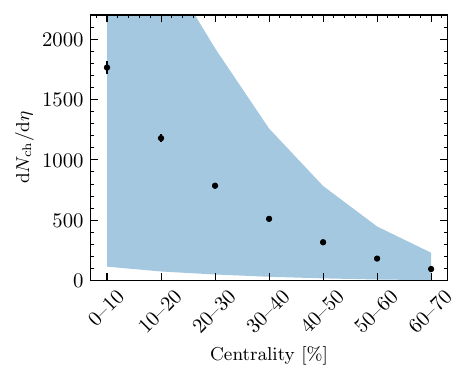}
		\includegraphics{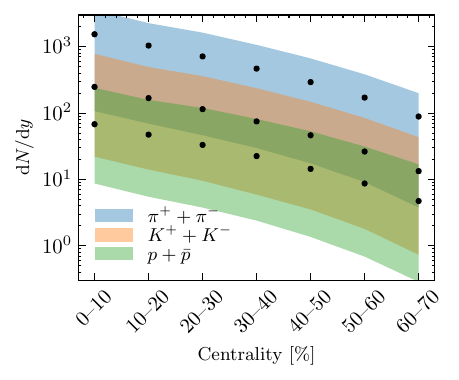}
		\includegraphics{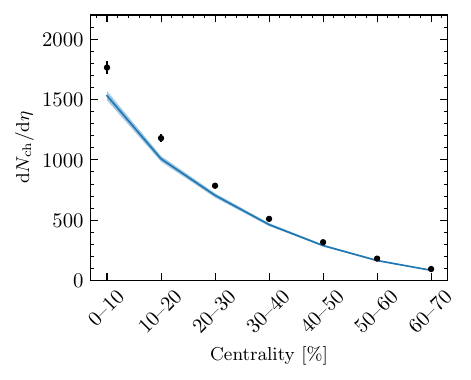}
		\includegraphics{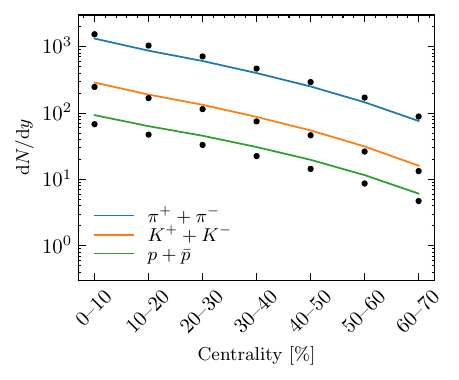}
		\caption{Top panel shows the model predictions obtained from parameter values sampled uniformly from the prior ranges for Pb+Pb collisions at $\sqrt{s_{NN}}=5.02$ TeV. Bottom panel compares the experimental data with model predictions evaluated at the median of the posterior parameter distribution in the analysis including spin polarization. The band represents the corresponding 1$\sigma$ posterior predictive uncertainty.}
		\label{fig:priorposterior_1}
	\end{figure*}

\begin{figure*}[t]
		\centering
		\includegraphics{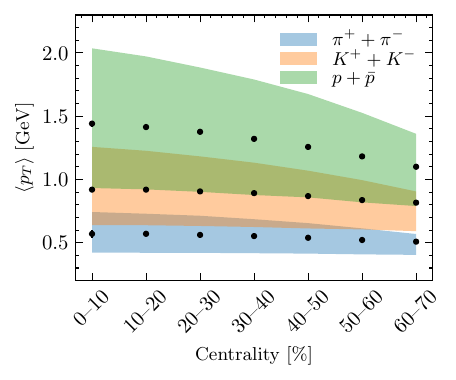}
		\includegraphics{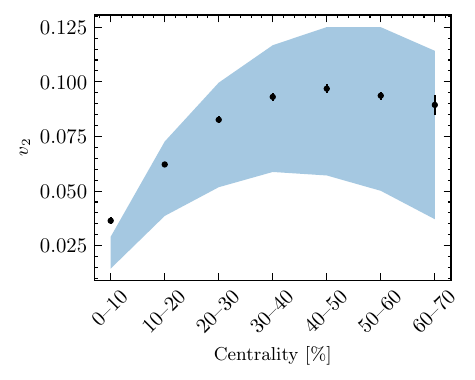}
		\includegraphics{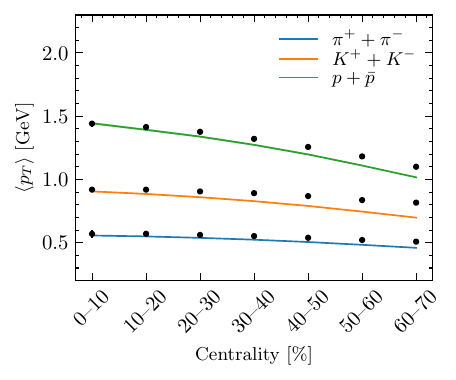}
		\includegraphics{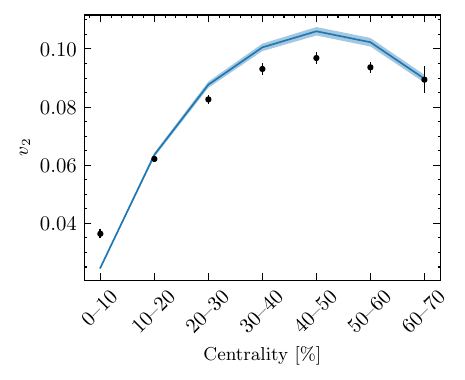}
		\caption{Top panel shows the model predictions obtained from parameter values sampled uniformly from the prior ranges for Pb+Pb collisions at $\sqrt{s_{NN}}=5.02$ TeV for $\langle p_T\rangle$ and $v_2$. Bottom panel compares the experimental data with model predictions evaluated at the median of the posterior parameter distribution in the analysis including spin polarization
		for $\langle p_T\rangle$ and $v_2$. The band represents the corresponding 1$\sigma$ posterior predictive uncertainty.}
		\label{fig:priorposterior_2}
    \end{figure*}

\begin{figure*}[t]
		\centering
		\includegraphics[width=0.49\textwidth]{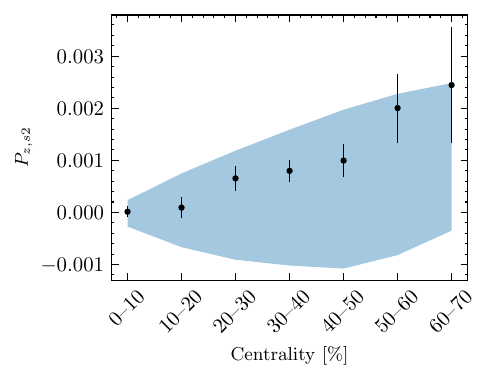}
		\includegraphics[width=0.49\textwidth]{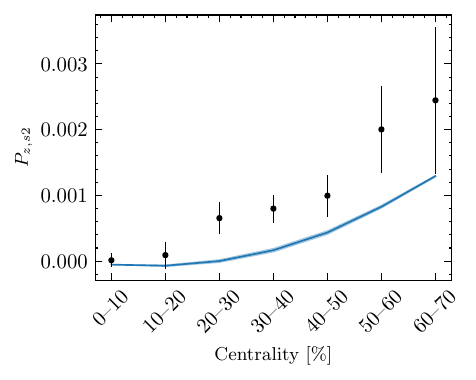}
		\caption{Left panel shows the model predictions obtained from parameter values sampled uniformly from the prior ranges for Pb+Pb collisions at $\sqrt{s_{NN}}=5.02$ TeV for longitudinal component of spin polarization.
			The right panel instead shows
			the experimental data with model predictions evaluated at the median of the posterior parameter distribution in the analysis including spin polarization for the longitudinal polarization.
		}
		\label{fig:priorposterior_3}
	\end{figure*}

	Specifically, the Latin hypercube samples are generated in the unit hypercube $[0,1]^d$ with $d=13$ using the \texttt{lhs} package in \textsc{R}~\cite{lhsR}, and are subsequently mapped to the physical model parameters by linearly scaling each dimension to its corresponding parameter range. This procedure corresponds to assuming independent uniform prior distributions for all model parameters within the specified bounds. The design consists of a total of 800 parameter sets.
	Fixed random seeds are used in the LHS construction to ensure reproducibility. We verify that the chosen prior ranges (listed in Table~\ref{tab::param_prior_map}) encompass the experimental data, as illustrated in Figs.~\ref{fig:priorposterior_1}-\ref{fig:priorposterior_3}. 
	
	\begin{table}[t]
		\caption{Prior range and MAP values of the parameters}
		\centering
		\begin{tabular}{|c|c|c|c|}
			\hline
			Parameter & Prior range & MAP value with $P_z$ & MAP value without $P_z$\\
			& (min--max)   & (rounded to two decimal places) & (rounded to two decimal places) \\
			\hline
			$\eta_0$  &  1.5--3.5    & 2.54  & 2.50\\
			$\sigma_\eta$ & 1.0--2.5  & 1.38 & 1.44\\
			$\eta_{B0}$  &  4.0--6.0  & 5.15 & 4.57\\
			$\sigma_{B,\text{in}}$ &  1.8--3.0  & 2.14 & 2.74\\
			$f$   &  0.0--0.4  &  0.17 & 0.16\\
			$\tau_0$ [fm]  &  0.6--1.2 &  0.83 & 1.0\\
			$(\eta/s)_{\text{min}}$ & 0.05--0.25 & 0.07 & 0.06\\
			$(\eta/s)_{\text{slope}}$ & 0.0--3.0 & 0.76 & 1.44\\
			$(\eta/s)_{\text{curv}}$ & -1.0--1.0 & 0.07 & 0.23\\
			$(\zeta/s)_{\text{max}}$ & 0.0--0.3 & 0.16 & 0.12\\
			$(\zeta/s)_{\text{width}}$ & 0.0--0.3 & 0.19 & 0.07\\
			$(\zeta/s)_{\text{T0}}$ & 0.1--0.3 & 0.16 & 0.15\\
			$\varepsilon_{\mathrm{sw}}$ [GeV fm$^{-3}$]   & 0.35--0.65 & 0.54 &  0.48\\
			\hline
		\end{tabular}
		\label{tab::param_prior_map}
	\end{table}
	
	The design points and their corresponding model outputs are stored in separate data files and are read into the analysis framework as input-output pairs. Prior to training the emulator, design points for which one or more observables are undefined or contain non-numerical values are removed from the dataset. In total, seven such design points are excluded, leaving $N_{\text{tot}}=793$ valid samples.
	
	The filtered dataset is divided into a training set and an independent validation set. A subset of 100 design points is randomly selected and reserved exclusively for validation, while the remaining samples are used for training the Gaussian process model emulator (discussed below). The random selection is performed using a fixed random seed to ensure reproducibility of the training-validation split. The training dataset (\textbf{X},\textbf{Y}) is used to construct the emulator, while the validation dataset (\textbf{X}$_\text{val}$,\textbf{Y}$_\text{val}$) is employed to test the emulator's accuracy and robustness.
	
	The model outputs are standardized to have zero mean and unit variance for each observable. This preprocessing is performed using a linear rescaling,
	$$Y_i\rightarrow \frac{Y_i-\mu_i}{\sigma_i}=Y_i^{\text{scaled}}$$
	where $\mu_i=\langle Y_i\rangle$ and $\sigma_i$ denote the sample mean and standard deviation of the $i^{\text{th}}$ observable across the training dataset. To reduce the dimensionality of the observable space and remove correlations among observables, we perform a principal component analysis (PCA) on the standardized model outputs using the \texttt{scikit-learn} Python library~\cite{JMLR:v12:pedregosa11a}. The full PCA spectrum is first computed to quantify the fraction of total variance carried by each principal component. We retain the leading $n_{\text{pc}}=3$ principal components, which together account for a 97.06\% of the total variance when performing analysis including spin polarization and 98.19\% when performing analysis excluding spin polarization. The remaining variance is incorporated into the likelihood function as truncation uncertainty, denoted as $\Sigma_{\text{trunc}}$, and therefore does not bias the parameter inference. Including additional principal components was found to degrade the emulator calibration, as the subleading modes are dominated by statistical noise and are not reliably constrained by the training data, leading to unstable Gaussian-process hyperparameters and spurious constraints in the Bayesian analysis.
	
	The PCA yields a transformation matrix \textbf{W} of dimension $n_{\text{pc}}\times n_{\text{obs}}$ which operates on standardized model outputs and projects them onto the retained principal components. In the present analysis $n_{\text{pc}}=3$,  resulting in a three-dimensional principal-component representation. The transformation reads
	$$\textbf{Z} = \textbf{W}\textbf{Y}^{\text{scaled}}$$
	The experimental observables are treated consistently within the same linear framwork. Let $\textbf{Y}_{\text{exp}}$ denote the vector of experimental mean values and let $\Sigma_{\text{exp}}$ denotes the covariance matrix of experimental uncertainties in observable space, assumed diagonal due to the absence of bin-to-bin correlations. The experimental observables are first standardized 
	$$\textbf{Y}_{\text{exp}}^{\text{scaled}} = \textbf{S}(\textbf{Y}_{\text{exp}}-\boldsymbol{\mu}),\quad \quad \quad \textbf{S}=\text{diag}(1/\sigma_i)$$
	Projection onto principal-component space yields
	$$\textbf{Z}_{\text{exp}} = \textbf{W}\textbf{Y}_{\text{exp}}^{\text{scaled}}$$
	and the corresponding covariance matrix transforms as
	$$\Sigma^{\text{PC}}_{\text{exp}} = \textbf{W}\ \textbf{S}\ \Sigma_{\text{exp}}\ \textbf{S}^\text{T}\  \textbf{W}^\text{T}$$
	The log-likelihood function used for Bayesian inference is given by
	$$\log \mathcal{L}(\theta) = -\frac{1}{2}\left[(\textbf{Z}-\textbf{Z}_{\text{exp}})^{T}\Sigma_{\text{tot}}^{-1}(\textbf{Z}-\textbf{Z}_{\text{exp}})+\log \text{det }\Sigma_{\text{tot}}\right] +\text{constant}$$
	where the total covariance in principal-component space is
	$$\Sigma_{\text{tot}}=\Sigma_{\text{emu}}+\Sigma^{\text{PC}}_{\text{exp}}+\Sigma_{\text{trunc}}$$
	with $\Sigma_{\text{emu}}$ denoting the emulator predictive covariance in principal-component space.
		
	\begin{figure*}[t]
		\centering
		\includegraphics{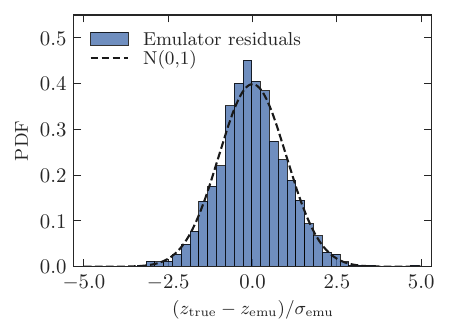}
		\includegraphics[width=\textwidth]{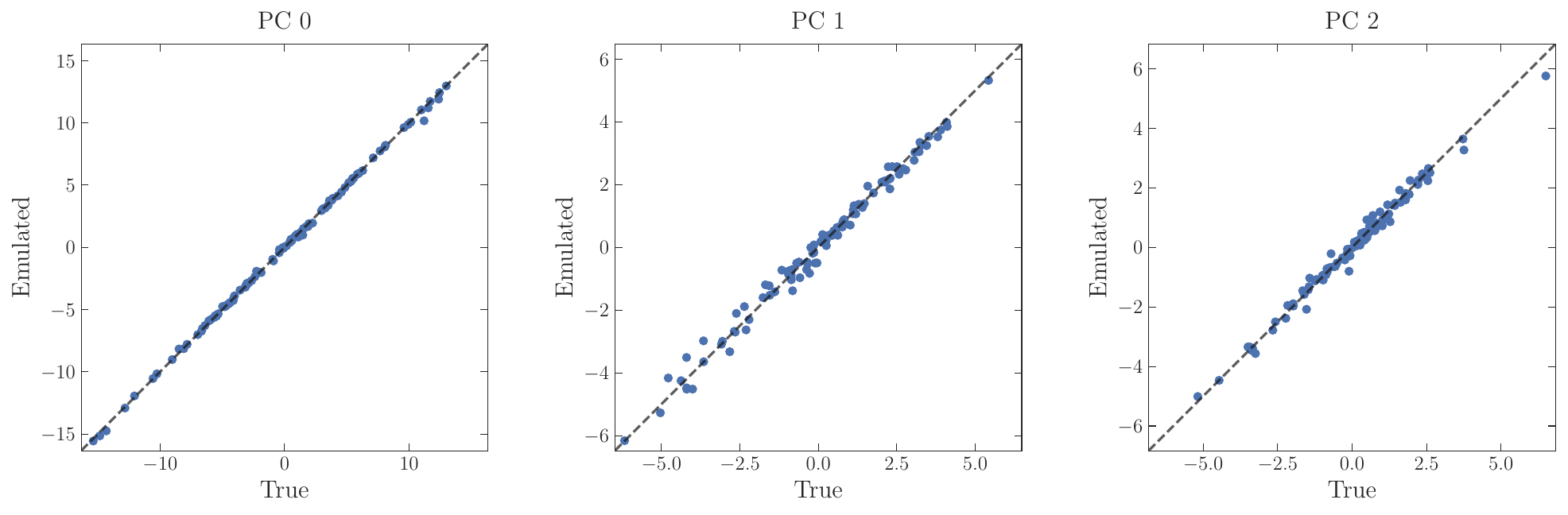}
		\caption{(Top) Distribution of normalized residuals ($Z$-scores) for the retained principal components. The dashed curve indicates a standard normal distribution. (Bottom) Comparison between the true model values and emulator predictions for the three principal components in the analysis including spin polarization. The diagonal line denotes perfect agreement.}
		\label{fig:emulator_test}
	\end{figure*}
	
	Each retained principal component is emulated independently using a Gaussian process regressor (GPR) from the \texttt{scikit-learn} library~\cite{JMLR:v12:pedregosa11a,gpy2014}. We employ a composite kernel consisting of a constant kernel multiplied by a squared-exponential (RBF) kernel, augmented by a white-noise term to account for residual numerical noise~\cite{Bernhard:2018hnz}. The characteristic length scales of the RBF kernel are initialized and bounded according to the physical ranges of the model parameters. Emulator performance is assessed using a train-test split of the design set. The coefficient of determination implemented as \texttt{GPR.score} in the \texttt{scikit-learn} library for the three retained princpal components is found to be 99.86\%, 96.88\%, and 95.97\%, respectively for the analysis with spin-polarization data. After confirming satisfactory predictive performance, the Gaussian process emulator is retrained using the full design dataset for the subsequent Bayesian analysis. The accuracy of the full emulator is further evaluated by reconstructing the observables at the validation design points (\textbf{X}$_\text{val}$,\textbf{Y}$_\text{val}$). Predictions obtained in principal-component space are mapped back to observable space using the inverse PCA transformation followed by inverse scaling. Emulator performance is quantified using the root-mean-square error (RMSE) in observable space defined as
	$$\text{RMSE} = \sqrt{\langle (Y_{\text{Emulated}}-Y_{\text{True}})^2\rangle}$$
	where $Y_{\text{True}}=Y_\text{val}$ denotes the model output. When normalized by the standard deviation of the validation data in observable space, we obtain an overall normalized RMSE of 3.5\%.

	To further assess emulator robustness, we perform a five-fold cross-validation of the Gaussian-process emulators in principal-component space. The design set is randomly partitioned into five subsets, and for each split the emulator is trained on four subsets and evaluated on the remaining one. For each principal component, we analyze the normalized residuals ($Z$-scores)~\cite{Hastie2009}, defined as the difference between the true and emulated values divided by the predicted emulator uncertainty. For a well-calibrated emulator, these residuals are expected to follow a standard normal distribution. Our results for the aggregated $Z$-score distribution is shown in the top panel of Fig.~\ref{fig:emulator_test} and is found to be consistent with a standard normal distribution. We additionally evaluate coverage probabilities, confirming that approximately 69.4\% and 95\% of the residuals fall within the corresponding one- and two-sigma confidence intervals. Finally, direct comparison between true and emulated principal-component values are performed to verify the absence of systematic bias as shown in the bottom panel of Fig.~\ref{fig:emulator_test}.
	
	\begin{figure*}[t]
		\centering
		\includegraphics{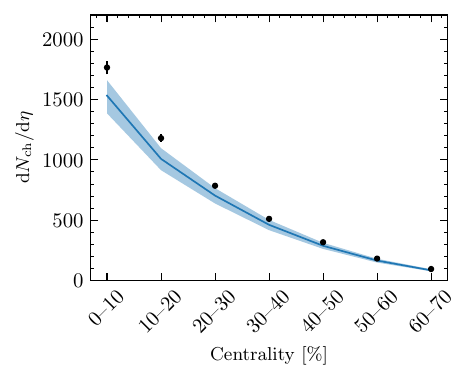}
		\includegraphics{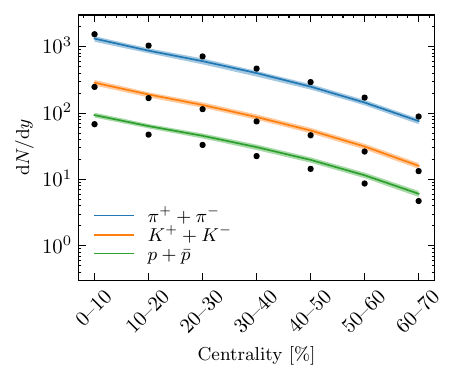}
		\includegraphics{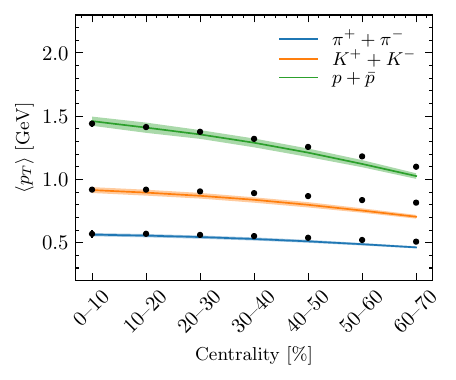}
		\includegraphics{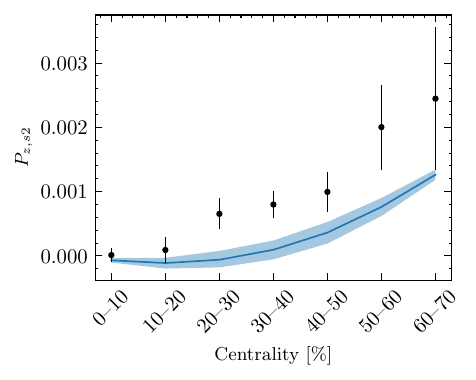}
		\includegraphics{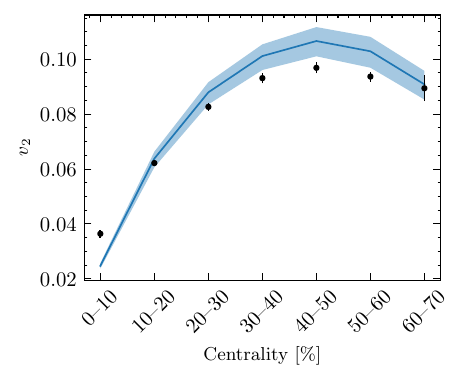}
		\caption{Posterior predictive checks for the Bayesian analysis of Pb+Pb collisions at $\sqrt{s_{NN}}=5.02$ TeV. The shaded bands correspond to the 5th–95th percentile intervals of the posterior predictive distribution obtained by propagating posterior parameter samples through the emulator trained including spin polarization.. The solid line indicates the median prediction. Experimental data are shown for comparison.}
		\label{fig:ppc}
	\end{figure*}
	
	The trained Gaussian-process emulators are used to predict the principal components for any parameter vector $\theta$, providing both the mean and diagonal emulator covariance. Predictions are mapped back to observable space via inverse PCA and inverse standardization, with emulator uncertainties propagated accordingly, including a truncation term for discarded variance, as discussed above. Uniform priors are assumed within the predefined parameter bounds, and the likelihood is defined as a multivariate Gaussian in principal-component space, incorporating emulator, experimental, and truncation covariances. Posterior distributions of the model parameters are sampled using the parallel-tempered MCMC algorithm implemented in \texttt{ptemcee} library~\cite{Foreman-Mackey_2013,10.1093/mnras/stv2422}, with multiple temperatures and walkers to ensure efficient exploration. After burn-in, the converged chains are used to infer the posterior distributions of all model parameters.
	
	\section{Results}
	\label{sec:results}
		
	\begin{figure*}[t]
		\centering
		\begin{minipage}{0.48\textwidth}
			\centering
			\hspace{6mm}\text{With $P_z$}\\[1mm]
			\includegraphics[width=\linewidth]{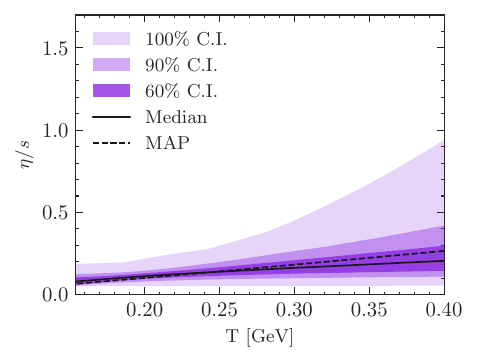}
		\end{minipage}
		\hfill
		\begin{minipage}{0.48\textwidth}
			\centering
			\hspace{8mm}\text{Without $P_z$}\\[1mm]
			\includegraphics[width=\linewidth]{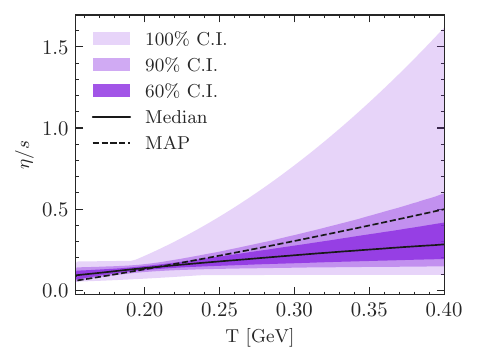}
		\end{minipage}
		\caption{Posterior distributions of the temperature-dependent shear viscosity obtained from the analysis with and without including $P_z$ in the left and right panels, respectively. The median estimate is shown as a solid line and the maximum-a-posteriori (MAP) estimate as a dashed line. The shaded bands correspond to the 60\%, 90\%, and 100\% credible intervals, with decreasing color intensity indicating increasing credibility level.}
		\label{fig:shear}
	\end{figure*}
	
	\begin{figure*}[t]
		\centering
		\centering
		\begin{minipage}{0.48\textwidth}
			\centering
			\hspace{6mm}\text{With $P_z$}\\[1mm]
			\includegraphics{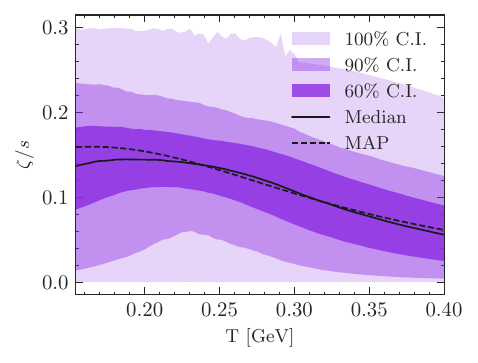}
		\end{minipage}
		\hfill
		\begin{minipage}{0.48\textwidth}
			\centering
			\hspace{8mm}\text{Without $P_z$}\\[1mm]
			\includegraphics{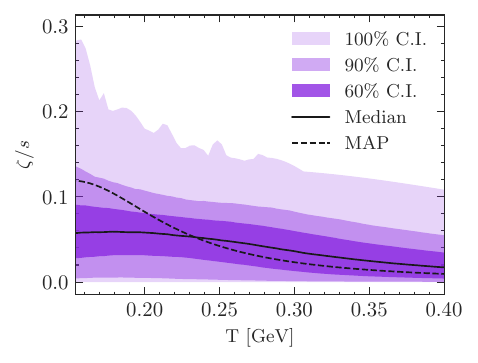}
		\end{minipage}
		\caption{Posterior distributions of the temperature-dependent bulk viscosity obtained from the analysis with and without including $P_z$ in the left and right panels, respectively. The median estimate is shown as a solid line and the maximum-a-posteriori (MAP) estimate as a dashed line. The shaded bands correspond to the 60\%, 90\%, and 100\% credible intervals, with decreasing color intensity indicating increasing credibility level.}
		\label{fig:bulk}
	\end{figure*}

	\begin{figure*}[t]
		\centering
		\begin{overpic}[scale=0.4]{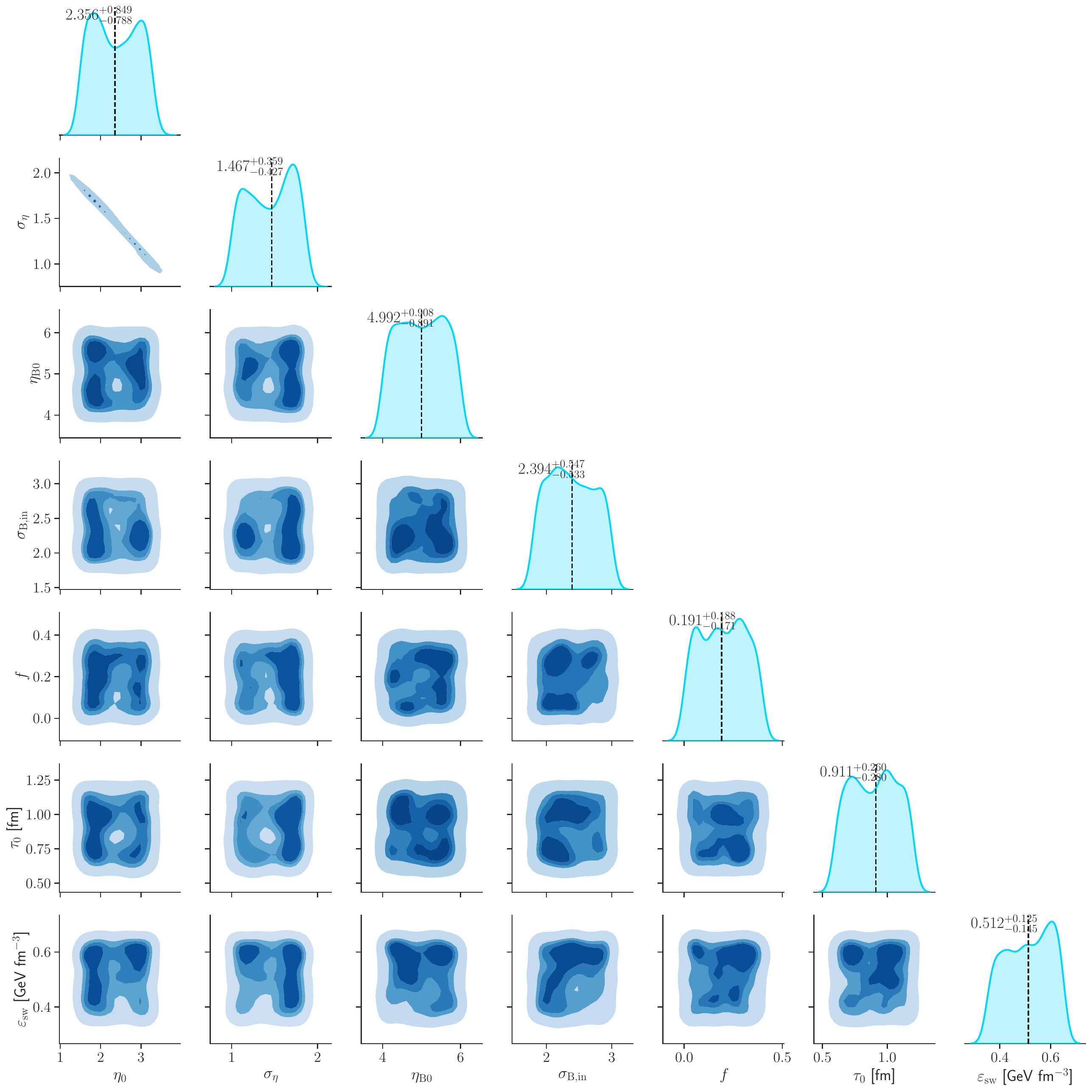}
			\put(50,90){\large \text{With $P_z$}}
		\end{overpic}
		\caption{Marginalized one- and two-dimensional posterior distributions of selected model parameters inferred from MCMC sampling using an emulator trained on bulk and spin polarization observable. The diagonal panels show one-dimensional marginal posteriors, while the off-diagonal panels display the corresponding two-dimensional joint posterior distributions. Dashed lines indicate posterior median values, and the quoted uncertainties represent central 90\% Bayesian credible intervals.}
		\label{fig:corner_ic_w}
	\end{figure*}
	
	\begin{figure*}[t]
		\centering
		\begin{overpic}[scale=0.4]{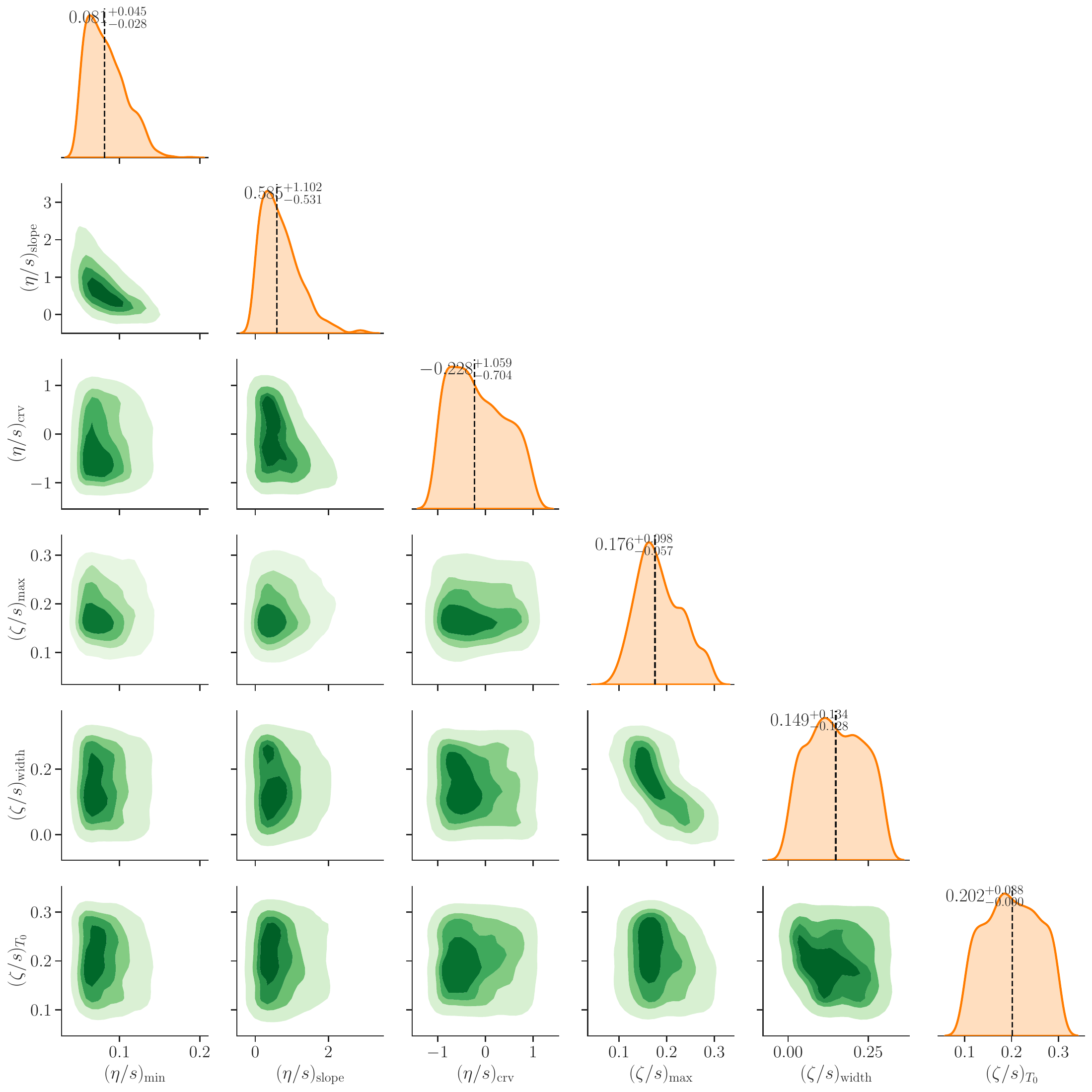}
			\put(50,90){\large \text{With $P_z$}}
		\end{overpic}
		\caption{Marginalized one- and two-dimensional posterior distributions of selected model parameters inferred from MCMC sampling using an emulator trained on bulk and spin polarization observable. The diagonal panels show one-dimensional marginal posteriors, while the off-diagonal panels display the corresponding two-dimensional joint posterior distributions. Dashed lines indicate posterior median values, and the quoted uncertainties represent central 90\% Bayesian credible intervals.}
		\label{fig:corner_trans_w}
	\end{figure*}
	
	\begin{figure*}[t]
		\centering
		\begin{overpic}[scale=0.4]{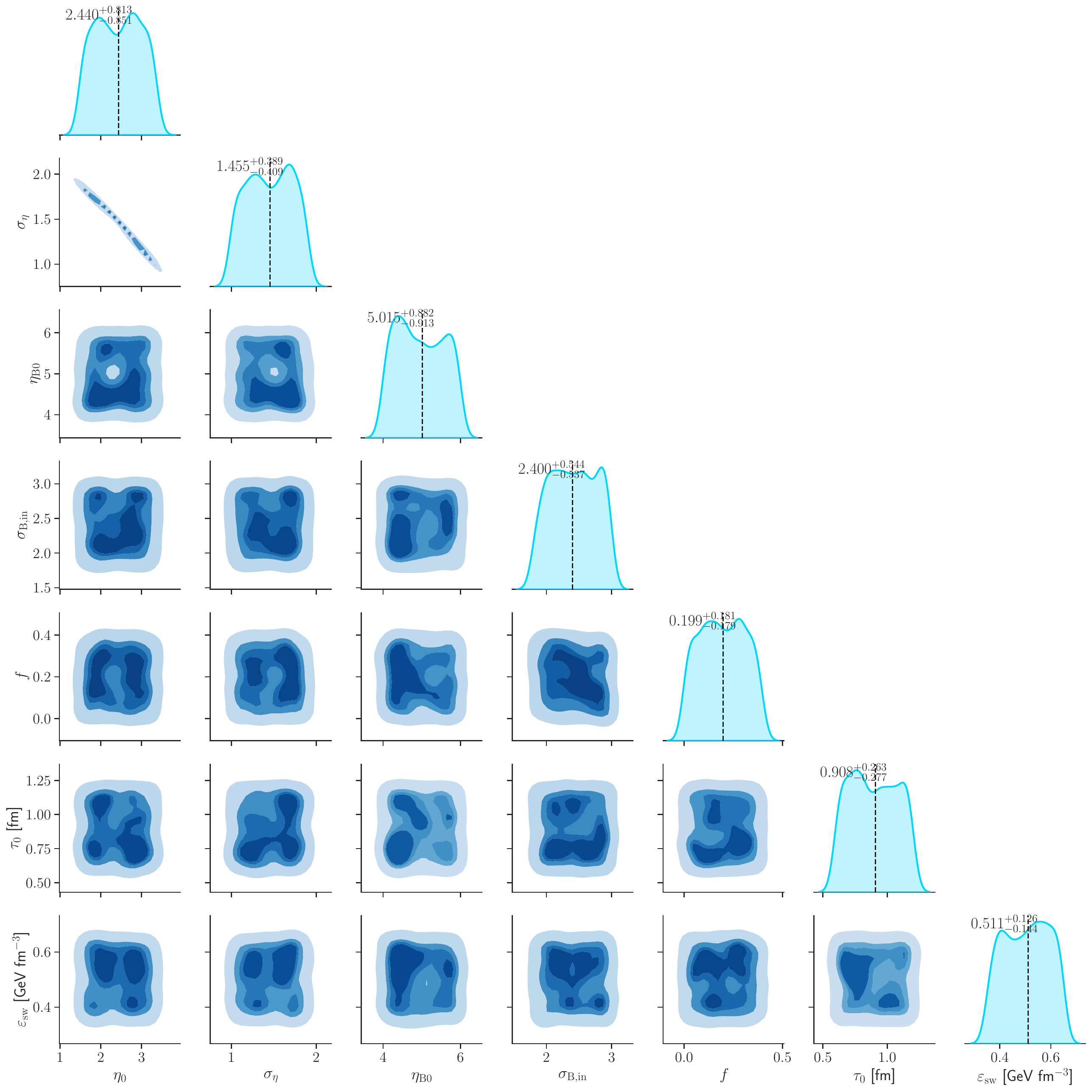}
			\put(50,90){\large \text{Without $P_z$}}
		\end{overpic}
		\caption{Marginalized one- and two-dimensional posterior distributions of selected model parameters inferred from MCMC sampling using an emulator trained on only bulk observables without spin. The diagonal panels show one-dimensional marginal posteriors, while the off-diagonal panels display the corresponding two-dimensional joint posterior distributions. Dashed lines indicate posterior median values, and the quoted uncertainties represent central 90\% Bayesian credible intervals.}
		\label{fig:corner_ic_wo}
	\end{figure*}
	
	\begin{figure*}[t]
		\centering
		\begin{overpic}[scale=0.4]{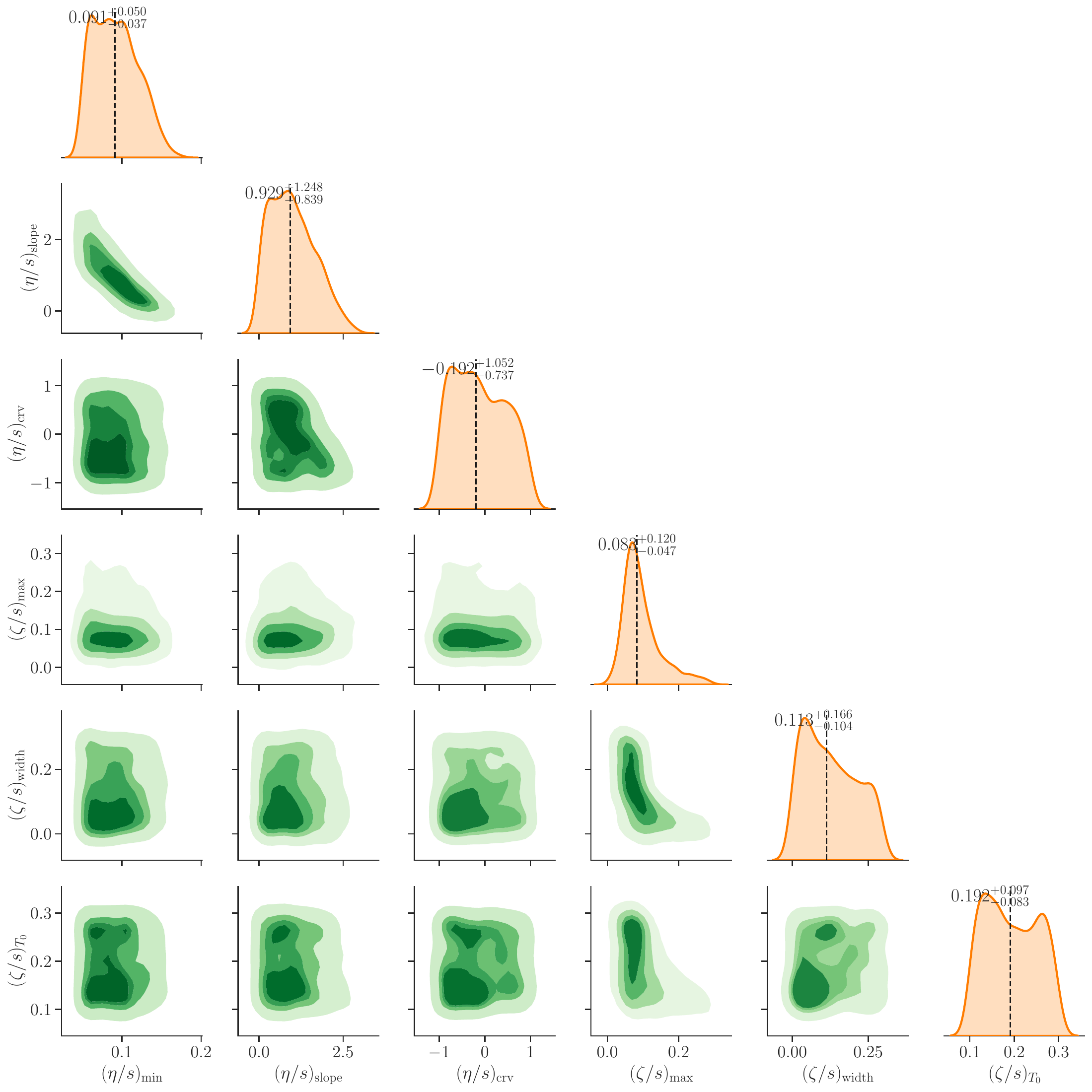}
			\put(50,90){\large \text{Without $P_z$}}
		\end{overpic}
		\caption{Marginalized one- and two-dimensional posterior distributions of selected model parameters inferred from MCMC sampling using an emulator trained on only bulk observables without spin. The diagonal panels show one-dimensional marginal posteriors, while the off-diagonal panels display the corresponding two-dimensional joint posterior distributions. Dashed lines indicate posterior median values, and the quoted uncertainties represent central 90\% Bayesian credible intervals.}
		\label{fig:corner_trans_wo}
	\end{figure*}
	
	To illustrate the outcome of the Bayesian parameter inference, we evaluate the emulator at the median of the posterior parameter distribution and compare the resulting model predictions with experimental data across all observables and centrality classes as shown in Figs.~\ref{fig:priorposterior_1}-\ref{fig:priorposterior_3}. The emulator provides the mean prediction in observable space together with the associated emulator uncertainty, obtained by propagating Gaussian-process uncertainties from principal-component space and including the PCA truncation variance. The shaded bands in the figures indicate the resulting 1$\sigma$ uncertainty on the model prediction. For comparison, we also show predictions obtained from parameter values sampled uniformly within the prior ranges, highlighting the degree to which the data constrain the model parameters. These comparisons demonstrate that the posterior-median parameter set yields a significantly improved description of the measured observables.
	
	To assess the overall consistency of the inferred model with experimental data, we perform posterior predictive checks. While the posterior-median comparison described above illustrates the predictions corresponding to a representative inferred parameter set, posterior predictive checks probe the full predictive distribution of the model and provide a consistency test of the Bayesian inference. Parameter samples are drawn from the posterior distribution and propagated through the trained emulator to generate replicated datasets in the observable space. The resulting ensemble of replicated observables is used to construct pointwise posterior predictive intervals, defined by the 5th and 95th percentiles, together with the median prediction. These intervals are compared directly to the experimental measurements across all observables and centrality classes, as shown in Fig.~\ref{fig:ppc}. Good agreement of the data with the posterior predictive bands indicates that the inferred parameter distributions provide a statistically consistent description of the measured observables. We emphasize that this procedure tests the full predictive distribution of the model rather than a single best-fit parameter set.

	Using the posterior samples obtained from the Bayesian analysis, we reconstruct the temperature dependence of the shear and bulk viscosity coefficients in Figs.~\ref{fig:shear}-\ref{fig:bulk}. For each coefficient, we evaluate the functions Eqs.~\ref{eq:etaS}-\ref{eq:zetaS} over a dense temperature grid using a large ensemble of posterior samples and construct pointwise credibility intervals. The median and maximum-a-posteriori (MAP) estimates are shown alongside the 60\%, 90\%, and 100\% credible intervals. We find that the extracted shear viscosity is essentially unchanged when longitudinal spin polarization data are included in the Bayesian analysis, indicating that bulk hadronic observables already provide strong constraints on shear transport. In contrast, the median value of the bulk viscosity is significantly enhanced-by approximately a factor of two-when spin polarization data are incorporated, although the MAP estimates remain comparable in the two analyses. Notably, both parameter analyses yield a satisfactory description of conventional bulk observables, suggesting that spin polarization provides complementary sensitivity primarily to bulk viscous effects rather than shear transport. This behavior is consistent with the expectation that longitudinal component of spin polarization is more sensitive to the space-time structure of the expansion dynamics of the medium, which are directly influenced by bulk viscosity.
	
	To visualize the inferred parameter correlations and the extent to which individual model parameters are constrained by the data, we construct corner (pairwise correlation) plots from the posterior samples, shown in Figs.~\ref{fig:corner_ic_w}--\ref{fig:corner_trans_wo}. For clarity, the parameters are separated into two groups: parameters associated with the initial condition model, initial hydrodynamic time $\tau_0$, and switching energy density $\varepsilon_{\mathrm{sw}}$; and parameters governing the temperature dependence of the shear and bulk viscosities. For each group, a random subset of posterior samples is used to generate kernel-density estimates of the one- and two-dimensional marginal distributions, with median values and central 90\% credible intervals indicated.
	
	We observe that parameters primarily related to the initial condition and hydrodynamic initialization exhibit nearly flat marginal distributions with broad, multi-modal structures and weak correlations. This indicates that, within the present analysis, these parameters are only weakly constrained by the considered experimental observables. Such behavior is expected, as the bulk hadronic observables and longitudinal spin polarization included here are predominantly sensitive to the medium's transport properties rather than to detailed features of the early-time dynamics or particlization hypersurface.
	
	\begin{figure*}[t]
		\centering
		\centering
		\begin{minipage}{0.48\textwidth}
			\centering
			\hspace{2.5cm}\text{$\pi^++\pi^-$}\\[1mm]
			\includegraphics{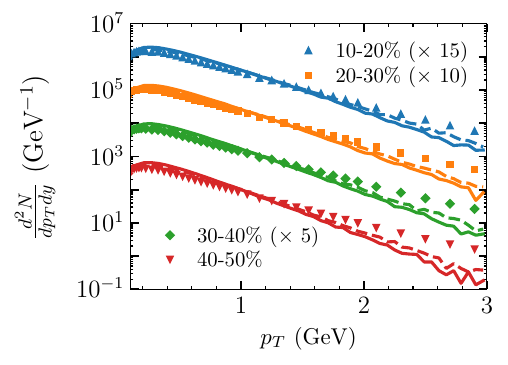}
		\end{minipage}
		\hfill
		\begin{minipage}{0.48\textwidth}
			\centering
			\hspace{2.5cm}\text{$K^++K^-$}\\[1mm]
			\includegraphics{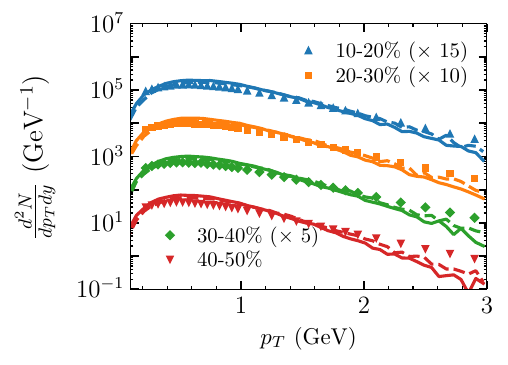}
		\end{minipage}
		\caption{Transverse momentum spectra of (left) charged pions and (right) charged kaons in Pb-Pb collisions at $\sqrt{s_{NN}}=5.02$ TeV for different centrality classes. Shown are model results obtained using median values of parameters extracted from Bayesian analyses with (solid lines) and without (dashed lines) inclusion of spin polarization observable. Experimental data is taken from Ref.~\cite{ExptData_identified_dNdy_5020}.}
		\label{fig:differential1}
	\end{figure*}
	
	In contrast, the posterior distributions of the parameters related to transport coefficients remain broadly consistent between the two analyses. The inclusion of longitudinal spin polarization does not lead to a substantial narrowing or broadening of the credible intervals, or to dramatic changes in the correlation structure among transport parameters. However, it does shift the inferred bulk-viscosity distribution toward larger values. In particular, the median of $\zeta/s(T)$ increases by approximately a factor of two when $P_z$ data are incorporated, while the MAP estimate exhibits a modest upward shift. The corresponding 68\% credible intervals retain significant overlap beyond $T\gtrsim 300$ MeV, indicating that the two analyses remain statistically compatible within uncertainties. These results suggest that longitudinal spin polarization provides complementary sensitivity to bulk viscous effects. In contrast, the extracted shear viscosity remains largely unchanged, reinforcing the conclusion that conventional bulk observables already provide strong constraints on shear transport.
	
	\begin{figure*}[t]
		\centering
		\centering
		\begin{minipage}{0.48\textwidth}
			\centering
			\hspace{2.5cm}\text{$p+\bar{p}$}\\[1mm]
			\includegraphics{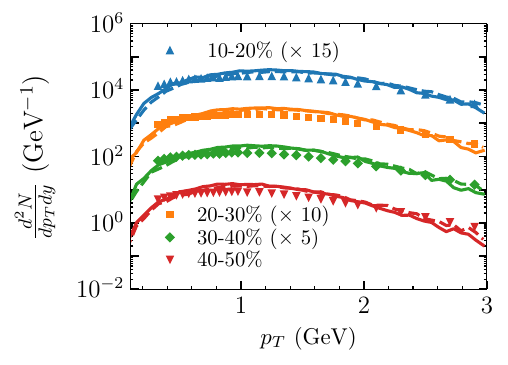}
		\end{minipage}
		\hfill
		\begin{minipage}{0.48\textwidth}
			\centering
			\hspace{2.5cm}\text{}\\[3mm]
			\includegraphics{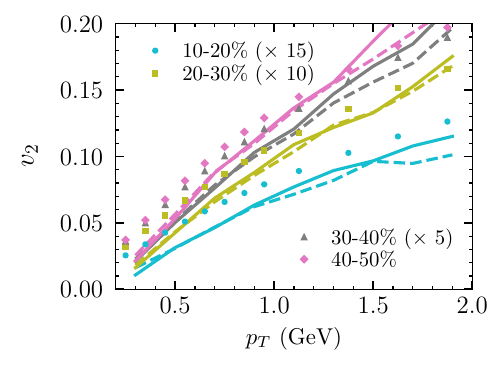}
		\end{minipage}
		\caption{Transverse momentum spectra of (left) protons and anti-protons and (right) the elliptic flow coefficient, $v_2$, of charged hadrons in Pb-Pb collisions at $\sqrt{s_{NN}}=5.02$ TeV for different centrality classes. The model results are obtained using the median parameter values extracted from Bayesian analyses with (solid lines) and without (dashed lines) the inclusion of spin polarization observable. The experimental data for the $p_T$-spectra is taken from Ref.~\cite{ExptData_identified_dNdy_5020}. The data for $v_2\{2\}$ and $v_2\{4\}$ from Ref.~\cite{ExptData_v2_5020} are used to construct mean $v_2$, as discussed in Section~\ref{sec:expdata}, and is shown in the right panel.}
		\label{fig:differential2}
	\end{figure*}

    \begin{figure*}[t]
    	\centering
    	\includegraphics[scale=0.95]{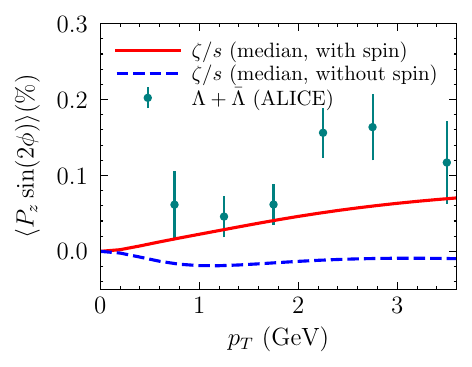}
    	\includegraphics[scale=0.95]{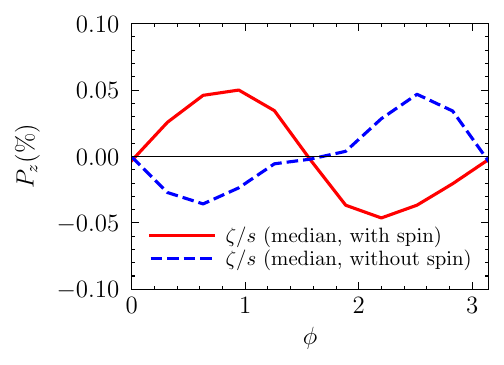}
    	\caption{(left) Transverse momentum and (right) azimuthal angle dependence of longitudinal polarization for Pb-Pb collisions at $\sqrt{s_{NN}}=5.02$ TeV in the 30-50\% centrality. Shown are model results obtained using median bulk viscosity values extracted from Bayesian analyses with and without inclusion of spin polarization observable.}
    	\label{fig:checkPz}
    \end{figure*}
	
	As an additional consistency check of our analysis, we use the median parameter values extracted from the corresponding posterior distributions to compute differential observables, including the transverse momentum spectra of pions, kaons, and protons, as well as the transverse momentum dependence of the charged-particle elliptic flow coefficient \(v_2\), shown in Figs.~\ref{fig:differential1} and \ref{fig:differential2}. We find that these differential hadronic observables are equally well described by both sets of extracted parameters, namely those obtained from analyses with and without spin polarization data. We additionally evaluate the transverse momentum and azimuthal angle dependence of the longitudinal spin polarization \(P_z\) shown in Fig.~\ref{fig:checkPz}. The results indicate that the experimentally observed sign of \(P_z\) is correctly reproduced only by the parameter set obtained from the Bayesian analysis that includes spin polarization observables. This parameter set corresponds to a bulk viscosity approximately a factor of two larger than that obtained in the analysis without spin observable. These findings further emphasize that spin polarization observables provide complementary and nontrivial constraints on the bulk viscosity of the QCD medium.
	
\begin{figure*}[t]
		\centering
		\centering
		\begin{minipage}{0.48\textwidth}
			\centering
			\includegraphics{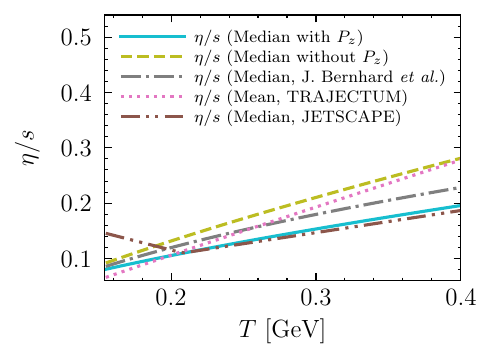}
		\end{minipage}
		\hfill
		\begin{minipage}{0.48\textwidth}
			\centering
			\includegraphics{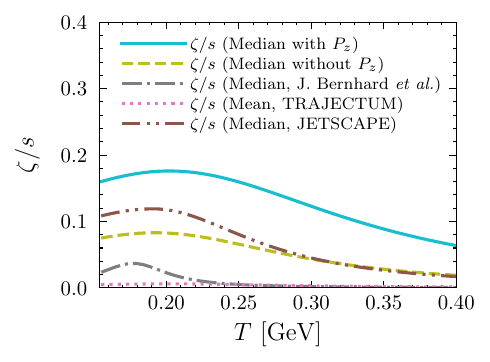}
		\end{minipage}
		\caption{Comparison of the extracted transport coefficients obtained from the present Bayesian analyses with and without spin observables (solid and dashed lines, respectively) with results from J.~Bernhard \emph{et al.}~\cite{Bernhard:2019bmu} (dash-dotted line), the TRAJECTUM framework~\cite{Trajectum2021} (dotted line), and the JETSCAPE collaboration~\cite{Everett:2021dsh} (dash-double-dotted line). The present results, the Bernhard \emph{et al.} analysis, and the JETSCAPE results employ median posterior parameter values, while the TRAJECTUM result corresponds to mean parameter values.}
		\label{fig:comparewothers}
	\end{figure*}

	Finally, we compare the extracted transport coefficients with results from other analyses in Fig.~\ref{fig:comparewothers}. The shear viscosity shown in the left panel is broadly consistent with existing results in the literature. In contrast, significantly larger variations are observed among the reported results for the bulk viscosity. The bulk viscosity obtained in our analysis without spin observable is in close agreement with the JETSCAPE result. However, the bulk viscosity extracted from the analysis including spin polarization observable is comparatively larger than existing results shown in the figure. Such a larger bulk viscosity appears to be necessary in order to reproduce the experimentally observed sign of the longitudinal polarization.
	
	\section{Summary}  
	In this work, we have performed a Bayesian analysis of Pb+Pb collisions at $\sqrt{s_{NN}}=5.02$ TeV using a Gaussian-process emulator trained on a finite ensemble of model evaluations. The predictive accuracy of the emulator was assessed by partitioning the design ensemble into training and validation subsets. The Gaussian-process emulator was trained on the training set and evaluated on the validation set, demonstrating nice agreement between predicted and true principal-component values. Posterior inference was carried out both with and without the inclusion of longitudinal spin polarization of $\Lambda$ hyperons in order to quantify its impact on the extraction of QGP transport properties.
	
	We find that the extracted shear viscosity $\eta/s(T)$ is largely unchanged when spin polarization data are included, indicating that conventional bulk hadronic observables already provide strong constraints on shear transport. In contrast, the inferred bulk viscosity $\zeta/s(T)$ exhibits an upward shift when $P_z$ is incorporated. In particular, the posterior median increases by approximately a factor of two, while the MAP estimate shows a modest increase. Although the credible intervals of the two analyses retain significant overlap, this shift indicates that longitudinal spin polarization provides additional sensitivity to bulk viscous effects.
	
	Overall, our results demonstrate that spin polarization measurements provide complementary information to conventional bulk observables in Bayesian extractions of QGP transport coefficients, particularly for bulk viscosity. These findings motivate more comprehensive future simulations and global analyses in which polarization observables are systematically incorporated alongside traditional hadronic measurements to further refine constraints on the dissipative properties of the quark–gluon plasma.
	
	\section{Acknowledgements}
	
	This work and S.K.S. were supported by ICSC - Centro Nazionale di Ricerca in High Performance Computing, Big Data 
	and Quantum Computing, funded by European Union - NextGenerationEU and by the project PRIN2022 Advanced Probes of 
	the Quark Gluon Plasma funded by ”Ministero dell’Università e della Ricerca”. 
	We further acknowledge the ICSC for awarding this project access to the EuroHPC supercomputer LEONARDO, hosted by 
	CINECA (Italy). S.K.S also thank Jhilam Sadhukhan for helpful discussions and assistance with debugging the 
	analysis script.
	\bibliography{refs}
\end{document}